\documentclass[12pt]{article} 
\oddsidemargin = \evensidemargin 
\newcommand{\D}{\displaystyle}
\usepackage{epsfig} 
\usepackage{colordvi} 
\usepackage{amssymb} 

\begin{document} 

\title{\bf Stability Analysis \\ of Coupled Map Lattices \\ 
at Locally Unstable Fixed Points} 

\author{Harald Atmanspacher$^{1,2}$, Thomas Filk$^{2,3}$, and Herbert Scheingraber$^1$ \\ \\ 
1 Center for Interdisciplinary Plasma Science, \\ 
Max-Planck-Institut f\"ur extraterrest\-rische Physik, \\ 85740 Garching, Germany \smallskip \\ 
2 Department of Theory and Data Analysis, \\ 
Institute for Frontier Areas of Psychology and Mental Health, \\ 
Wilhelmstr.~3a, 79098 Freiburg, Germany 
\smallskip \\ 
3 Institute of Physics, University of Freiburg, \\ 
Hermann-Herder-Str.~3, 79104 Freiburg, Germany} 

\date{} 
\maketitle 

\bigskip
\centerline{Accepted for publication in {\it European Physical Journal B}}

\bigskip
\centerline{PACS: \ 05.45.Ra,  \ 05.45.Xt, \ 89.75 Hc}

\vskip 1.5cm 

\begin{abstract} 

Numerical simulations of coupled map lattices (CMLs) and other complex model systems
show an enormous phenomenological variety that is difficult to classify and understand. 
It is therefore desirable to establish analytical tools for exploring fundamental
features of CMLs, such as their stability properties. Since CMLs can be considered as 
graphs, we apply methods of spectral graph theory to analyze their stability at locally
unstable fixed points for 
different updating rules, different coupling scenarios, and different types of neighborhoods. 
Numerical studies are found to be in excellent agreement with our theoretical results. 
\end{abstract} 


\vfill\eject

\section{Introduction} 

Coupled map lattices (CMLs) are arrays of cells whose state value is continuous, 
usually within the unit interval, over discrete space and time. Starting with 
Turing's seminal work on morphogenesis \cite{turi52}, they have 
been used to study the behavior of complex spatiotemporal systems for more than 
50 years. More recently, Kaneko and collaborators have established many 
interesting results for CMLs \cite{kane93} as generalizations of 
cellular automata, whose state values are discrete.     

One key motivation for modeling spatiotemporally extended systems with CMLs 
is to simplify the standard approach in terms of partial differential 
equations. And, of course, CMLs would not have become accessible without 
the rapid development of fast computers with large storage capacities. 
Within the last decades, CMLs have been applied to the study of areas as diverse 
as social systems, ecosystems, neural networks, spin lattices, Josephson junctions, 
multimode lasers, hydrodynamical turbulence, and others (cf.~the special 
journal issues {\it CHAOS} {\bf 2}(3), 1992, and {\it Physica D} {\bf 103}, 1997).        

A compact characterization of a CML with one time parameter is given by 
\begin{equation} 
\label{eq:maincml}
  u(n+1,x) = (1-\epsilon) f(u(n,x)) + \frac{\epsilon}{n_x} 
     \sum_{y\sim x} g(u(n,y)) ~, 
\end{equation} 
where $x$ represents the sites of the lattice, $x=1, ..., N_{tot}$,
here considered as the vertices of a graph, and $n$ represents the time 
step of the iteration. The parameter $\epsilon$ 
specifies the coupling between each cell and its neighborhood (and 
is often considered as constant over time and space). 
The sum over $y\sim x$ is the sum over 
all $n_x$ neighbors $y$ of vertex $x$. The function $g$ characterizes the interaction
of a vertex with its neighborhood and will be explained below.  

As in many 
other studies of CMLs, $f(x)$ is the logistic map on the unit interval, 
\[           f(x) = r x (1-x)  \, , \] 
with $0\le r \le 4$. For $r\ge 1$ the logistic map has a critical point at 
${r-1}\over r$ which is unstable for $3 < r \le 4$. 
The relevance of maps with quadratic maximum (such as the logistic map) 
for models of neurobiological networks was recently substantiated 
by novel results concerning a non-monotonic (rather than sigmoid) 
transfer function for individual neurons \cite{kuhn04}. 

For $\epsilon\rightarrow 0$, there is no coupling at all; hence, local 
neighborhoods have no influence on the behavior of the CML.  This situation 
represents the limiting case of $N_{tot}$ independently operating local objects 
at each lattice 
site. In the general case $0< \epsilon <1$,  the independence of individual cells 
is lost and the lattice behavior is governed by both local and 
global influences, depending on the chosen neighborhood. 
CMLs with a maximal neighborhood, $n_x \approx N_{tot}$, 
are often denoted as globally coupled maps.     
Their behavior is determined by global properties alone (mean field 
approach). 

The function   
\begin{equation}
\label{eq:g}
   g(x) = \alpha x + \beta f(x) + \gamma f(f(x)) ~, 
\end{equation} 
with 
\[   \alpha + \beta + \gamma = 1  \hspace{1cm} 
       \alpha , \beta , \gamma  \geq 0 ~, \] 
allows us to treat the interaction between each vertex and its 
neighborhood in different ways, depending on its time scale $\Delta t$. 
If the interaction can be regarded as simultaneous, $\Delta t \approx 0$, 
the situation can be approximated by $\alpha = \gamma = 0$ and 
$\beta =1$. Such a type of coupling, sometimes called ``future coupling'' 
\cite{meht00}, will be referred to as {\it non-causal coupling} 
\cite{atma04a} in the following, since the simultaneity of the interaction between 
vertex and neighbors makes a clear distinction of cause and effect 
impossible.     

The situation of a finite interaction time $\Delta t > 0$ can be properly 
modeled by $\beta = \gamma = 0$ and $\alpha =1$. In this way, past states 
in the neighborhood of a vertex are considered to act on the present 
state of the vertex with limited signal speed, so that the effect of 
an interaction is delayed with respect to its cause. Such a type of 
coupling will therefore by denoted as {\it causal coupling} in the 
following. Corresponding lattice behavior has recently been studied in 
\cite{meht00, maso03, liet04, atay04, atma04b}.

A third, somewhat exotic possibility arises for $\alpha = \beta = 0$ 
and $\gamma =1$. This case reflects the idea to model the action of future states 
of a vertex neighborhood on a present vertex state. More precisely, this 
refers to {\it locally extrapolated} future states and is justified for small
$\epsilon$ since then $u(n+1,y) \approx f(u(n,y))$. In this interpretation,  
the case of non-vanishing $\gamma$ is in 
contradiction with causality; thus we refer to such a situation 
as {\it anti-causal coupling}. 

The inclusion of anti-causal coupling can 
be interesting if one wants to study the consequences of a decomposition 
of a fundamental time-reversal invariant 
evolution of a system into forward and backward components. 
Using such a decomposition, one can investigate the influence of both
components on the stability properties of CMLs. It has indeed been found
that the stability of CMLs is supported by large $\alpha$, and their
stability is obstructed if $\alpha$ is small \cite{atma04a}. Based
on these observations, one may speculate that stability acts as a selection 
criterion for a forward arrow of time \cite{atma05}.   

Another time scale important for the physical interpretation of eq.~(1) is 
the time interval $\Delta \tau$ assumed for the updating mechanism, i.e.~for 
the physical integration of signals from the neighborhood states with the 
vertex state considered. If signals between vertices are 
transmitted much slower than the time scale assumed for the updating mechanism, 
$\Delta \tau \ll \Delta t$, the updating can be implemented (almost) instantaneously, 
or {\it synchronously}. If this is not the case, $\Delta \tau \gtrsim \Delta t$, 
updating must be implemented {\it asynchronously}. This entails the 
additional problem of determining a proper updating sequence, which can be 
random or depend on particular features of the situation considered. 

Most of the work on CMLs 
published in this respect (cf.~\cite{kane00}) was based on synchronous 
updating. For asynchronous updating as, for instance, studied by Lumer \& Nicolis 
\cite{lume94}, it was found that the behavior of CMLs differs strongly from that of CMLs with 
synchronous updating. Additional results for asynchronous updating were reported by 
\cite{marc97, rolf98, meht00, atma04a, atma04b}. 
Asynchronous updating rules have been suggested as particularly relevant for 
neurobiological networks. 

As a common feature of the (so far) few studies of asynchronous updating, it has 
been reported that it facilitates the synchronization and stabilization of CMLs 
decisively. In particular, Mehta \& Sinha \cite{meht00} demonstrated that the dynamics 
at individual lattice cells is strongly synchronized by coupling among cells. 
Atmanspacher \& Scheingraber \cite{atma04a, atma04b} showed that unstable 
fixed points at individual vertices can be stabilized as 
a consequence of their coupling to neighboring unstable fixed points.      

Such a stabilization is of particular interest since it is independent of external 
control mechanisms. The global stabilization of unstable local behavior operates 
inherently, without external adjustment, 
once the coupling is strong enough. Such a possibility represents a 
powerful alternative to external control procedures in the style of ``controlling 
chaos'' \cite{ott90}. 

Most published results concerning the stabilization and synchronization of CMLs
have been obtained by numerical studies for various parameters due to different 
coupling, different neighborhoods, and different updating. A number of theoretical 
approaches have been developed to describe these numerical results analytically
\cite{mack95,loss95,lema96,lema98,bely00}. 
In this contribution we present, for the first time, a comprehensive 
stability analysis of CMLs at locally unstable fixed points, which allows us 
to understand corresponding numerical results theoretically. 
In particular, the sophisticated dependence of the stabilization behavior 
on the different parameters mentioned above \cite{atma04a, atma04b} 
can be clarified. 

As recently proposed by Jost \& Joy \cite{jost02} and by Belykh et al.~\cite{bely04}, 
the stability properties of CMLs can be compactly and conveniently analyzed in a 
graph theoretical manner. We first determine, in section \ref{sec:spectrum}, the 
eigenvalues of the adjacency matrices related to various 
neighborhoods on regular lattices. In section \ref{sec:stabil}, 
we derive a stability condition for synchronous update and constant solutions of
the logistic map and apply it to special cases for causal and non-causal coupling 
numerically studied in \cite{atma04a, atma04b}. In addition, 
theoretical and numerical results are obtained for the situation of anti-causal coupling.
In all cases, the agreement between theoretical and numerical results is excellent.
In section \ref{sec:asynch}, we present an
intuitive argument for a stability analysis with asynchronous update.
In section \ref{sec:general}, the discussion is extended to generalized
neighborhoods.
Section \ref{sec:summary} summarizes and concludes the paper, 
and some perspectives are addressed. 

\section{The Spectrum of a Graph}
\label{sec:spectrum}

A CML can be considered within the more general framework
of the theory of graphs, whose properties can be studied by well-developed 
analytic methods \cite{bigg74, cvet95}. A graph is defined by
a set $F$ of vertices, a set $E$ of edges, and a 
map $\delta^{\pm}: E\rightarrow F$ indicating whether a vertex is an initial
point or a terminal point for an edge. A CML can be implemented on a graph with 
no self-loops, no multiple edges, and only undirected edges. Such graphs 
are sometimes referred to as {\it simple} graphs (cf.~\cite{wils85}). Many interesting properties of 
CMLs, in particular their stability properties, can then be investigated by two basic
matrices characterizing the corresponding graph: its {\it adjacency matrix}
and its {\it valence matrix}.      

For an arbitrary simple graph (defined by the ``neighbor-relation'' 
$\sim$ between its vertices), the adjacency matrix $A(x,y)$ and
the valence matrix $V(x,y)$ are defined by:
\[   A(x,y) = \left\{ \begin{array}{ll}
      1 & \mbox{for }y \sim x \\ 0 & {\rm otherwise} 
             \end{array} \right.  \]
and
\[  V(x,y) = \left\{ \begin{array}{ll}
      n_x & \mbox{for } x=y \\
      0 & \mbox{otherwise} \end{array}
        \right. 
     \hspace{1cm}
    {\rm with}~~
    n_x = \sum_y A(x,y) ~, \]
where $n_x$ is the number of neighbors of vertex $x$. The matrix $A$  
contains only 1s and 0s (no multiple edges), is non-reflexive (zero diagonal
since there are no self-loops) and symmetric (undirected edges).  
If the valence matrix is proportional
to the identity matrix, $V = v\,{\bf 1}$, the graph is called {\it regular}.


For the eigenvalues $\lambda$ of $V^{-1}A$ we have $|\lambda| \le 1$
since $V^{-1}A$ is a Markov matrix, i.e.~$(V^{-1}A)_{ij} \ge 0$ and
$\sum_j (V^{-1}A)_{ij} = 1$. The eigenvalue
$\lambda_{\rm max}=+1$ is assumed for the constant 
vector $u(x)$. However, we will see that the lower
bound for the eigenvalues of $V^{-1}A$ can deviate remarkably
from $-1$. From ${\rm tr}\,V^{-1}A=0$ we
can only deduce that the lower bound has to be negative (for infinite
graphs it can be zero). 

In the following, we consider regular graphs, $V=v\,{\bf 1}$.
We determine the eigenvalues for the adjacency matrix
for different types of neighborhoods on a 2-dimensional 
$N\times N$ square lattice with periodic boundary conditions. In particular, we focus on the
following types of neighborhoods: von Neumann 1st order (the four nearest
neighbors horizontally and vertically), Moore 1st order (the eight nearest 
neighbors including the diagonal neighbors), von Neumann 2nd order (Moore 1st
order plus the next to nearest neighbors horizontally and vertically), 
and Moore 2nd order (the 24 lattice sites in a
$5\times 5$ square around the central point). 

The eigenvalues are calculated by introducing
the shift matrix:
\[     P = \left( \begin{array}{ccccccc}
    0 & 1 & 0 & 0 & 0 & \cdots & 0 \\
    0 & 0 & 1 & 0 & 0 & \cdots & 0 \\
   \vdots & \vdots & \vdots & \vdots & \vdots & \vdots \\
    0 & 0 & 0 & 0 & 0 & \cdots & 1 \\
    1 & 0 & 0 & 0 & 0 & \cdots & 0 \end{array} \right) ~.\]
$P$ is a unitary matrix with $P^N={\bf 1}$ and
$P^{-1}=P^+ = P^{N-1}$. Therefore, the eigenvalues
$\omega_k$ of $P$ satisfy $\omega_k^N=1$ and are given by
the $N$ square roots of 1:
\[     \omega_k = \exp \left( {\rm i} \phi \right) = \exp \left( \frac{2\pi {\rm i}k}{N}\right) 
       \hspace{1cm} {\rm with~} k=0,1,\ldots,N-1 ~. \]
On a 2-dimensional $N\times N$ lattice 
we define the two shift operators for each direction 
by:
\[    P_1 = P \otimes {\bf 1} \hspace{1cm}
     {\rm and} ~~~~ P_2 = {\bf 1} \otimes P ~. \]
The adjacency matrices for the different neighborhoods commute
with $P_1$ and $P_2$ and can be completely expressed in terms of
these operators:
\begin{eqnarray*}
  A_{\rm v.\,N.\,1} &=& P_1 + P_1^{-1} + P_2 + P_2^{-1}  \\
  A_{\rm M.\,1} &=& ({\bf 1}+P_1+P_1^{-1})({\bf 1}+P_2+P_2^{-1}) - {\bf 1} \\
  A_{\rm v.\,N.\,2} &=& ({\bf 1}+P_1+P_1^{-1})({\bf 1}+P_2+P_2^{-1}) - {\bf 1}
    + P_1^2 + P_1^{-2} + P_2^2 + P_2^{-2}  \\
  A_{\rm M.\,2} &=& ({\bf 1}+P_1+P_1^{-1} + P_1^2 + P_1^{-2} )
      ({\bf 1}+P_2+P_2^{-1} + P_2^2 + P_2^{-2}) - {\bf 1}  ~.
\end{eqnarray*}
Therefore, the eigenvalues of $A$ are given by:
\begin{eqnarray*}
  \mbox{v.\,N.\,1}: & &   2\cos \phi_1 + 2 \cos \phi_2 \\
   & &  \hspace{1.5cm}{\rm with}~~ \phi_i = \frac{2\pi k_i}{N} 
         ~~{\rm where}~k_i = 0,1,\ldots N-1 ~.  \\
  \mbox{M.\,1}: & & (1+2\cos \phi_1)(1+2\cos \phi_2) - 1 \\
  \mbox{v.\,N.\,2}: & & (1+2\cos \phi_1)(1+2\cos \phi_2) - 1 
     + 2 \cos 2 \phi_1 + 2 \cos 2\phi_2 \\
  \mbox{M.\,2}: & & (1+2\cos \phi_1+ 2\cos 2 \phi_1)
    (1+2\cos \phi_2+2\cos 2 \phi_2) - 1 \\
\end{eqnarray*}
A discrete Fourier ansatz
\[
u_{k_1,k_2}(x_1,x_2) = \exp \left( {2\pi i \over N} (k_1 x_1 + k_2 x_2) \right)
\]
yields the same eigenvalues.
Setting $k_1=k_2=0$, we verify that the maximal eigenvalue of $A$ is 
always equal to the number of neighbors.

Assuming a large lattice size $N$, the minimal eigenvalues are
determined by differentiating with respect to $\phi_i$.
For the range of eigenvalues $\lambda$ of $V^{-1}A$ we finally obtain:
\begin{eqnarray*}
  \mbox{v.\,N.\,1}: & &  - 1 \leq \lambda_{k} \leq 1 \\ 
  \mbox{M.\,1}: & &  - \frac{1}{2} \leq \lambda_{k} \leq 1 \\
  \mbox{v.\,N.\,2}: & & - \frac{13}{36} \leq \lambda_{k} \leq 1 \\
  \mbox{M.\,2}: & &  - \frac{29}{96} < \lambda_{k} \leq 1 ~.
\end{eqnarray*}
(In general, the lower bound given in these expressions refers to the continuous
limit $N\rightarrow \infty$ of $\phi$ and is only approximately realized for discrete
$\phi$, i.e.~finite $N$.)  
For the case of global coupling one finds $-\frac{1}{N^2 - 1} \le \lambda \le 1$
and, in the limit $N\rightarrow\infty$, $0\le \lambda \le 1$.

The spectrum of a regular graph can be similarly obtained for higher 
dimensions as well as for other lattice types. As an example we mention 
a triangular lattice (six nearest neighbors)
which can be considered as a square lattice with the diagonals
in one direction added. The adjacency matrix is given by
\[   A_{\rm triangle} = P_1 + P_1^{-1} + P_2 + P_2^{-1} + P_1 P_2 +
           P_1^{-1} P_2^{-1} ~, \]
and the corresponding eigenvalues of $V^{-1}A$ are:
\[   \lambda_{k_1,k_2} = \frac{1}{6} \left( 2 \cos \phi_1 + 2 \cos \phi_2
       + 2 \cos (\phi_1 + \phi_2) \right)  ~. \]
The maximal eigenvalue is $\lambda_{\rm max}=+1$ and the minimal
eigenvalue is $\lambda_{\rm min}= -\frac{1}{2}$.  
This is identical to the case of a square lattice with a 1st order Moore 
neighborhood. 

\section{Stability Analysis for Constant Solutions with Synchronous Updating}
\label{sec:stabil}

Now we consider the stability of a constant solution,
\[    \bar{u}(n,x) \equiv \bar{u} ~~~{\rm for~all}~
      n, x~,  \]
of a CML (eq.\ \ref{eq:maincml}).
Given our normalization of parameters, an example of such a constant solution
is the critical point of the logistic map: $\bar{u} = \frac{r-1}{r}$. (As will be shown
in Sec.~3.3, this is -- apart from $\bar u = 0$ -- the only constant solution if $\gamma = 0$.) 
Furthermore, we 
consider a sychronous update of the lattice, which strictly corresponds to 
eq.\ (\ref{eq:maincml}). Special features related to asynchronous
updating will be discussed in section \ref{sec:asynch}.

Let $\{u_k(x)\}$ be a complete set of solutions of the eigenvalue
equation for the normalized adjacency matrix $V^{-1}A$:
\[    \sum_y (V^{-1}A) \, (x,y) u_k(y) = \frac{1}{n_x} \sum_{y\sim x} u_k(y) 
          = \lambda^k u_k(x)~. \]
For the stability analysis we consider fluctuations around
the constant solution,
\[   u_n(x) = \bar{u} + \delta a^k_n u_k(x) ~, \]
and obtain:
\begin{equation}
\label{eq:delta}
  \delta a^k_{n+1} = (1-\epsilon) f'(\bar{u}) \delta a^k_n
   + \epsilon \lambda_k g'(\bar{u}) \delta a^k_n ~.
\end{equation}
Hence, the constant solution is stable if $\epsilon$ satisfies
the following inequalities:
\begin{equation}
\label{eq:stab1}
   -1 < (1-\epsilon) f'(\bar{u})
      + \epsilon \lambda_k g'(\bar{u}) < +1 
   \hspace{0.6cm} \mbox{for all } k ~. 
\end{equation}
Inserting the logistic map for $f$, and using $g$ as defined in
eq.\ (\ref{eq:g}) we find:
\[    f'(\bar{u})= 2-r  \hspace{1cm} {\rm and} ~~~
     g'(\bar{u})= \alpha - (r-2) \beta + (r-2)^2 \gamma ~. \]
This leads to the stability conditions:
\begin{equation}
\label{eq:stabil}
  r-3 < \epsilon \left( r-2 + \lambda_k [\alpha
      - (r-2)\beta + (r-2)^2 \gamma] \right) <  r-1 
    \hspace{1cm} \mbox{for all } k~. 
\end{equation}
Note that $\alpha - (r-2)\beta + (r-2)^2\gamma$ can be positive
or negative. Depending on this sign, it is the smallest or the
largest eigenvalue $\lambda_k$ of the normalized adjacency matrix for
which one of the inequality conditions in eq.\ (\ref{eq:stabil}) is first  
violated. 

Due to the many parameters ($\alpha$, $\beta$, $\gamma$, $\epsilon$ and $r$) 
in the CML (eq.\ \ref{eq:maincml}), and in the stability conditions
(eq.\ \ref{eq:stabil}), we restrict our discussion to special
cases emphasizing those parameter ranges which have
been numerically investigated in previous work. 


\subsection{Pure Causal Coupling} 

Pure causal coupling is characterized by $\beta=\gamma=0$ and
$\alpha=1$. The stability conditions (eq.\ \ref{eq:stabil})
then read:
\[   r-3 < \epsilon \left( r-2 + \lambda_k \right)
     < r-1 ~. \]
The lower inequality defines the critical coupling strength:
\begin{equation}
\label{eq:eps}
   \epsilon_{\rm c} = 
       \frac{r-3}{r-2 + \lambda_{\rm min}} ~. 
\end{equation}
Inserting the different minimal eigenvalues for the 
various neighborhoods we find:
\begin{eqnarray*}
   \mbox{v.\,N.\,1  } & & \epsilon_{\rm c} =
            \frac{r-3}{r-2-1} = 1  \\  
   \mbox{M.\,1  } & & \epsilon_{\rm c} =
            \frac{r-3}{r-2.5}   \\  
   \mbox{v.\,N.\,2  } & & \epsilon_{\rm c} =
            \frac{r-3}{r-2-\frac{13}{36}} =  \frac{r-3}{r-2.361...}    \\  
   \mbox{M.\,2  } & & \epsilon_{\rm c} =
            \frac{r-3}{r-2-\frac{29}{96}} = \frac{r-3}{r-2.302...}  \\
   \mbox{global} & & \epsilon_{\rm c} =  
            \frac{r-3}{r-2}~.
\end{eqnarray*}

These results can be compared with numerical simulations presented in
\cite{atma04b}. Figure 1
gives a compact representation of those results, showing the critical
coupling strength $\epsilon_c$, beyond which the unstable fixed point of $u$
is stabilized, as a function of the control parameter $r$  
of the logistic map for different neighborhoods as indicated in the figure and for synchronous updating. 
The case of global coupling turns out to be realized for
asynchronous updating (cf.~Sec.~4) as well.  

Note that for the first order von Neumann neighborhood 
the constant solution is always unstable for $\epsilon < 1$.  
For $r=4$ (right hand side of Fig.~1) we obtain the 
following critical values of $\epsilon$:
\begin{eqnarray*}
   \mbox{v.\,N.\,1  } & & \epsilon_{\rm c} = 1  \\  
   \mbox{M.\,1  } & & \epsilon_{\rm c} = 0.66... \\
   \mbox{v.\,N.\,2  } & & \epsilon_{\rm c} = 0.61... \\  
   \mbox{M.\,2  } & & \epsilon_{\rm c} = \frac{96}{163} = 0.589...  \\
   \mbox{global} & & \epsilon_{\rm c} = 0.5~.
\end{eqnarray*}

It is easy to check that the numerical results of Fig.~1 agree perfectly 
well with the theoretical results according to eq.~(6).     

\medskip
\renewcommand{\baselinestretch}{0.85}
\begin{figure}[htb] 
\begin{center} 
\vskip 0.2cm 
\epsfig{figure=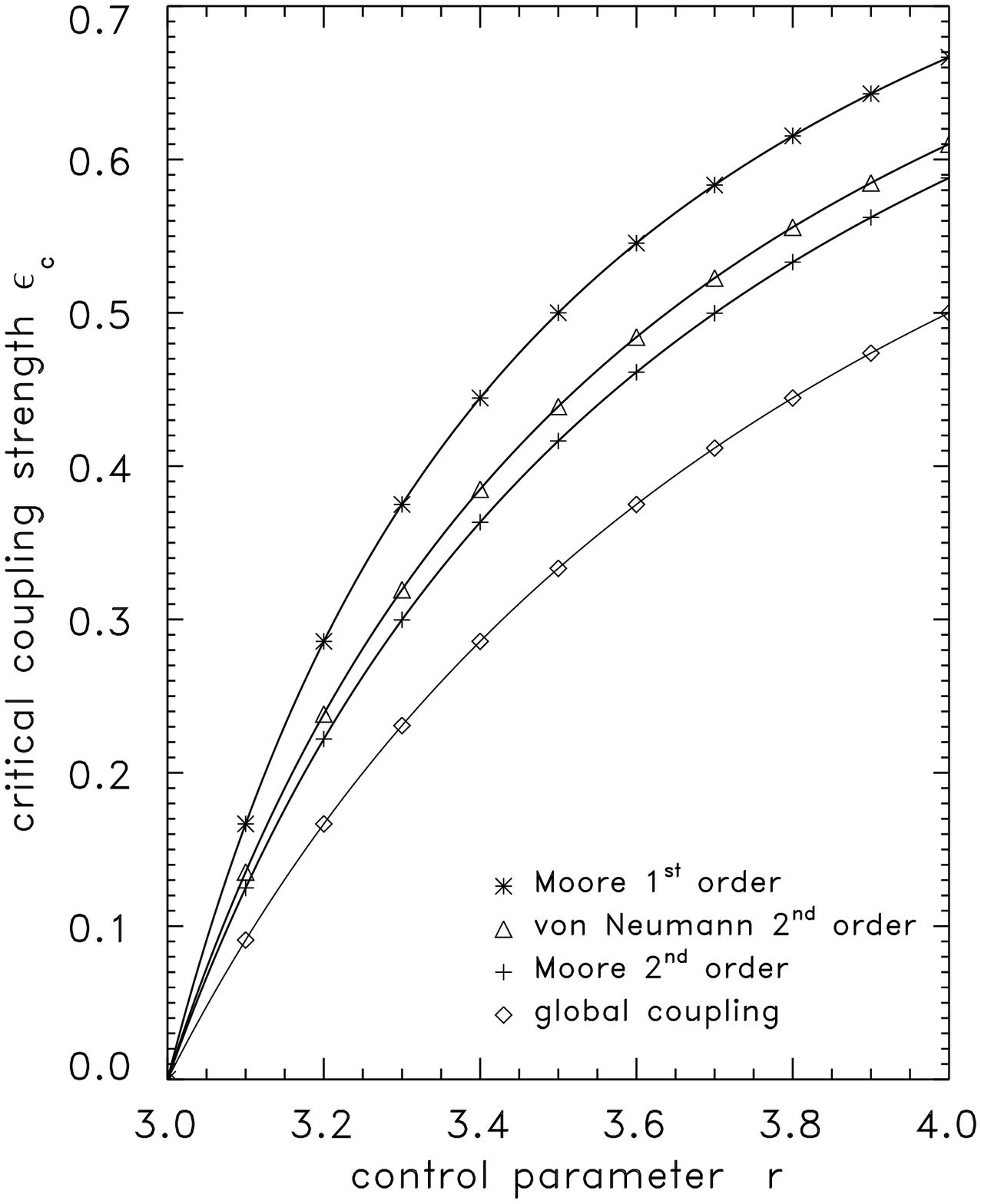,scale=0.5} 
\end{center} 
\label{fig1}
\begin{quote} 
{\footnotesize Figure 1: Critical coupling strength for stabilization onset, $\epsilon_c$, 
as a function of $r$, $3 \le r \le 4$. 
The individual values are numerical results for synchronous 
updating. All curves are given by the functions of $\epsilon_c$ on $r$ according
to eq.~(6).}     
\end{quote} 
\end{figure} 

\subsection{Causal and Non-Causal Coupling} 

Next we consider the case $\gamma=0$, $\beta=1-\alpha$, and
$r=4$. This situation has been investigated numerically in
\cite{atma04a}. From eq.~(5), we now
obtain the stability conditions:
\begin{equation}
\label{eq:stabilnoncausal}
    1 < \epsilon \left( 2 + \lambda_k (3 \alpha - 2)
                                             \right) < 3  ~. 
\end{equation}

For $\alpha = \frac{2}{3}$, this becomes 
independent of the neighborhood. In this case the critical value 
$\epsilon_c$ derived from the lower bound of eq.~(7) is:
\[    \epsilon_{\rm c}(\alpha={ \textstyle \frac{2}{3}})
      = \frac{1}{2}  ~. \]

For $\alpha> \frac{2}{3}$, the minimal eigenvalue of the
adjacency matrix determines the stability of the CML with respect to
the lower bound in eq.~(7) and we find:
\begin{eqnarray*}
   \mbox{v.\,N.\,1  } & & \epsilon_{\rm c} = \frac{1}{4-3\alpha} \\  
   \mbox{M.\,1  } & & \epsilon_{\rm c} =
            \frac{1}{3-\frac{3}{2}\alpha}   \\  
   \mbox{v.\,N.\,2  } & & \epsilon_{\rm c} =
            \frac{36}{98-39\alpha}   \\  
   \mbox{M.\,2  } & & \epsilon_{\rm c} =
            \frac{96}{250-87\alpha}  \\
   \mbox{global } & & \epsilon_{\rm c} =
            \frac{1}{2} ~.
\end{eqnarray*}

For $\alpha<\frac{2}{3}$, the lower bound
is first saturated for the maximal eigenvalue $\lambda_{\rm max}=1$ 
of the normalized adjacency matrix. Since this is independent of the
neighborhood and holds for any simple graph, we have the result:
\[        \epsilon_{\rm c} = \frac{1}{3\alpha} ~. \]

As $\epsilon \leq 1$ there is no critical value for $\epsilon$
for $\alpha < \frac{1}{3}$. In this regime the constant solution is generally 
unstable (for synchronous updating). 
 
Figure 2 shows
the numerical results for combinations of causal and non-causal coupling
for both synchronous and asynchronous updating. All results for synchronous
updating agree perfectly with our theoretical results. Asynchronous updating
will be discussed separately in Sec.~4.

\renewcommand{\baselinestretch}{0.85}
\begin{figure}[htb] 
\begin{center} 
\vskip 0.2cm 
\epsfig{figure=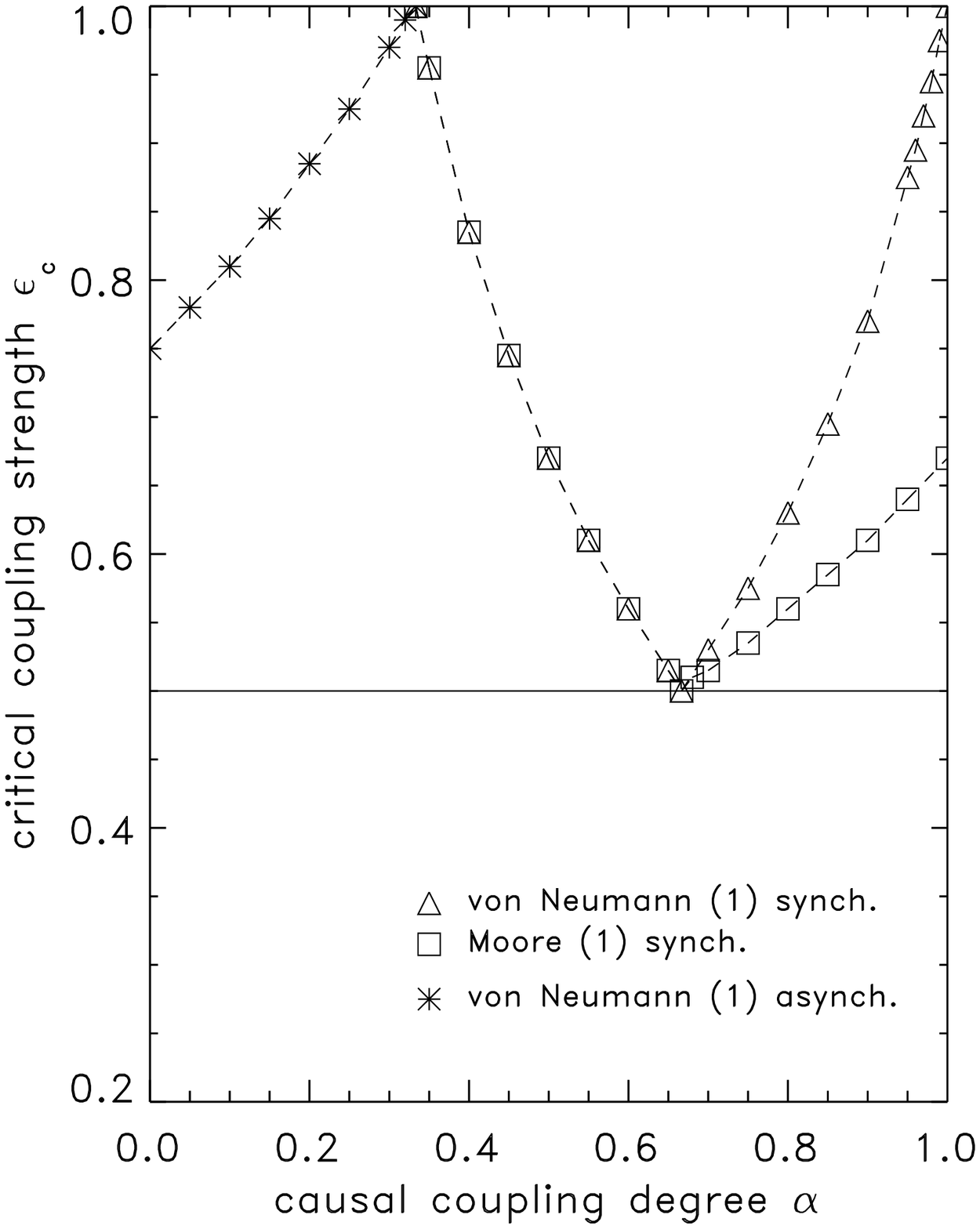,scale=0.55} 
\end{center}
\label{fig2} 
\begin{quote} {\footnotesize Figure 2: Critical coupling strength $\epsilon_{c}$ 
for stabilization onset as a function of the degree $\alpha$ of causal coupling 
for $r = 4$. Different symbols refer to different 
neighborhoods as explained in the figure. The solid line at $\epsilon_c = 0.5$
characterizes the situation for asynchronous updating, independent of the neighborhood
considered.}     
\end{quote} 
\end{figure} 

\subsection{Causal, Non-Causal, and Anti-Causal Coupling}

This case may be of less practical interest but can shed some light on 
general features of CMLs. In particular, it presents features which
are not shared by the coupling scenarios discussed in the preceding sections. 
We distinguish three cases: pure anti-causal coupling 
($\gamma=1$, $\alpha=\beta=0$), combinations of causal and
anti-causal coupling  ($\beta=0$, $\alpha=1-\gamma$), and
combinations of non-causal and anti-causal coupling 
($\alpha=0$, $\beta=1-\gamma$). Our discussion refers mostly to the stability of
the constant solution at $\bar u = \frac{r-1}{r}$. For $\gamma \neq 0$ there are
additional constant solutions which depend on $\epsilon$.

\subsubsection{Pure Anti-Causal Coupling: $\gamma=1$, $\alpha=\beta=0$}

The stability conditions in this case are given by:
\[  r-3 < \epsilon \left( r-2 + \lambda_k (r-2)^2
     \right) < r-1 ~. \]
For the lower bound, we only obtain solutions in the
range $0 \leq \epsilon \leq 1$ if:
\begin{equation}
\label{eq:cond}
 \lambda_{\min} > - \frac{1}{(r-2)^2} ~. 
\end{equation}

For $r=4$, this condition is not fulfilled for the 
neighborhoods considered in this paper except for global coupling.

For a given graph, eq.\ (\ref{eq:cond}) is satisfied if 
\[  r  \leq 2 + \sqrt{\frac{1}{|\lambda_{\rm min}|}} ~. \]
In this case, the stability conditions are given by
\[   \frac{r-3}{(r-2) - |\lambda_{\rm min}|
       (r-2)^2}  ~<~  \epsilon ~<~
     \frac{r-1}{(r-2) + (r-2)^2}  ~. \]   

\subsubsection{Causal and Anti-Causal Coupling: $\beta=0$, $\alpha=1-\gamma$}

For $r=4$, the stability conditions now assume the form:
\[  1 < \epsilon \left( 2 + \lambda_k (1+3\gamma)
     \right) < 3   ~. \]
There is a general upper bound
\[     \epsilon < \frac{1}{1+\gamma} ~, \]
beyond which the constant solution at the unstable fixed point remains unstable. 
The lower bound for $\epsilon$ depends on the type of neighborhood:
\[
\begin{array}{lcll}
   \mbox{v.\,N.\,1  } & & \epsilon >
      \D   \frac{1}{1-3\gamma}  \hspace{1cm} & 
                 \mbox{nowhere satisfied}  \\[0.3cm]  
   \mbox{M.\,1  } & & \epsilon > \D
            \frac{2}{3-3\gamma}  & \mbox{if} \ \ \D \gamma < \frac{1}{3} \\[0.3cm]  
   \mbox{v.\,N.\,2  } & & \epsilon > \D
            \frac{36}{59-39\gamma}  & \mbox{if} \ \ \D \gamma < \frac{23}{39} \\[0.3cm]  
   \mbox{M.\,2  } & & \epsilon > \D
            \frac{96}{163-87 \gamma} & \mbox{if} \ \ \D \gamma < \frac{67}{87} \\[0.3cm]
 \mbox{global  } & & \epsilon > \D
            \frac{1}{2} & \mbox{for all} \ \gamma ~.
\end{array}
\]

\renewcommand{\baselinestretch}{0.85}
\begin{figure}[h] 
\begin{center} 
\epsfig{figure=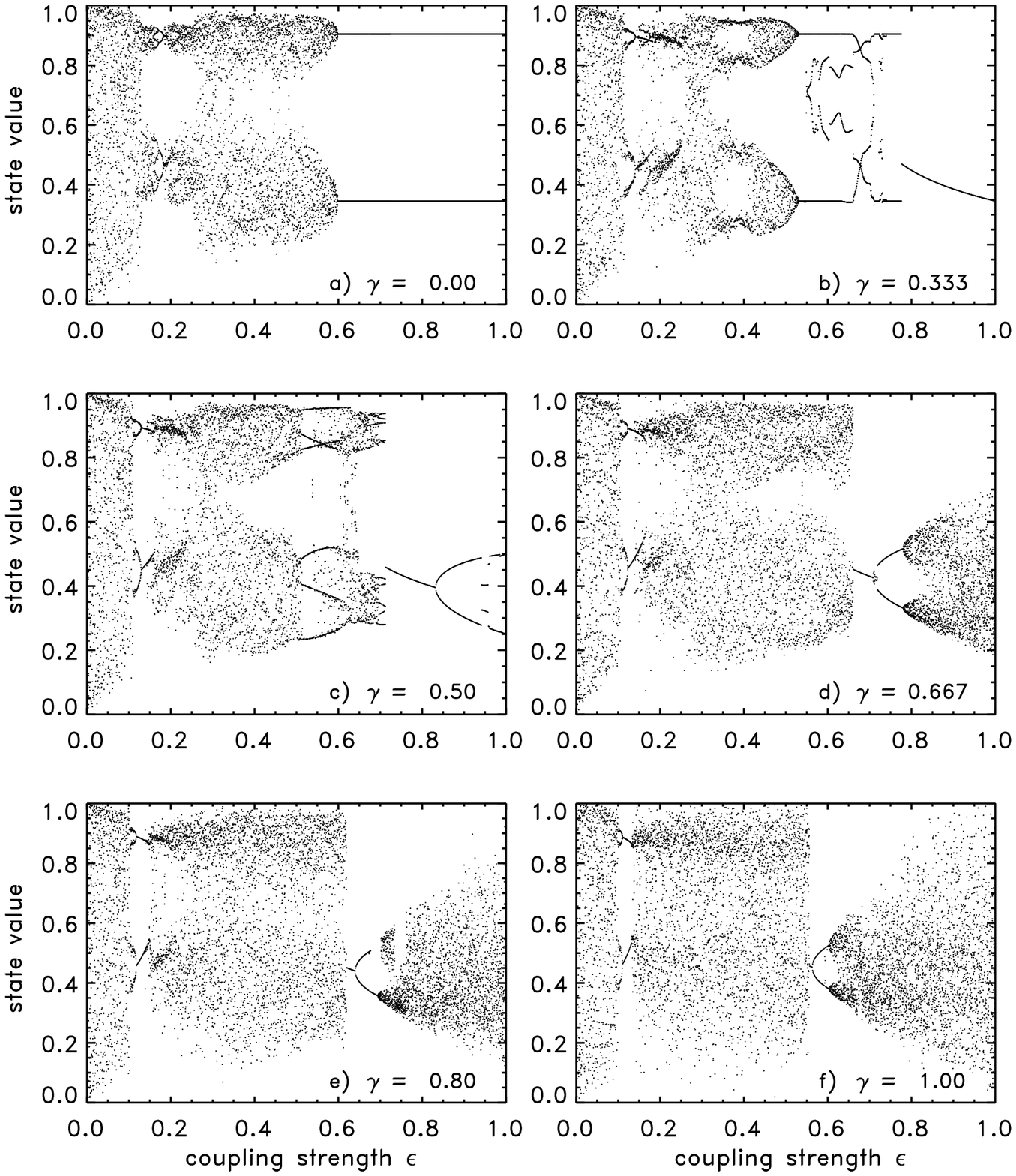,scale=0.7} 
\end{center}
\label{fig2} 
\vspace{-0.2cm} 
\begin{quote} {\footnotesize Figure 3: Stability diagram for synchronously updated CMLs 
with a von Neumann neighborhood of order 1. 
For selected values of $\gamma=1-\alpha$, the distribution of states
is plotted versus coupling strength $\epsilon$. Simulations are based on random initial
conditions on a $50 \times 50$ lattice, at least 5000 iterations, and $r=4$.}     
\end{quote} 
\end{figure}

Fig.~3 shows numerically obtained CML state values as a function of coupling 
strength $\epsilon$ for selected values of $\gamma$ and for a first order von 
Neumann neighborhood. There is no stable solution at the unstable
fixed point, corresponding to our theoretical result that the stability condition
is nowhere satisfied in this case. 
However, we observe $\epsilon$-dependent solutions different from the
unstable fixed point. This is due to the fact that for constant solutions 
and $\gamma \neq 0$ eq.~(1) is of fourth order. The solutions can be 
characterized (for $r=4$) by \begin{equation}
\bar{u}_{1/2} = \frac{5}{8} \pm \frac{1}{8}
     \sqrt{ 5 - \frac{4(1-\epsilon)}{\epsilon \gamma}} ~. 
\end{equation}
Real solutions exist only for $\epsilon \geq \frac{4}{4+5\gamma}$ as 
can be recognized by the stable regimes in Fig.~3b-e. 
In Fig.~3a there is only a bistable solution.  

\subsubsection{Non-Causal and Anti-Causal Coupling: $\alpha=0$, $\beta=1-\gamma$}

For several interesting values of $\gamma$, Fig.~4 shows numerically obtained state values of 
the CML as a function of coupling strength $\epsilon$ for a first order von Neumann neighborhood,
Fig.~5 shows the same for a first order Moore neighborhood.

\renewcommand{\baselinestretch}{0.85}
\begin{figure}[h] 
\begin{center} 
\epsfig{figure=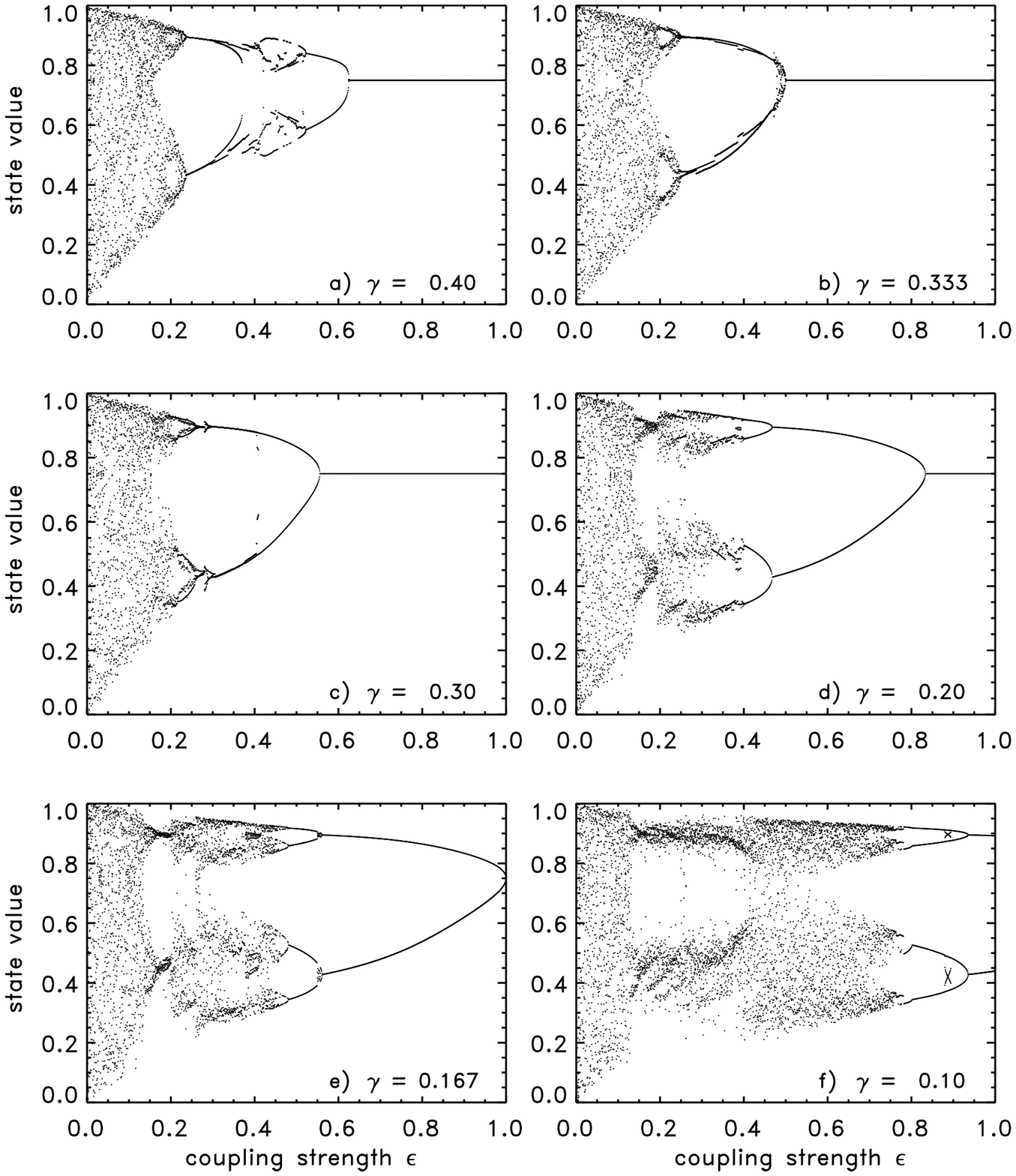,scale=0.7} 
\end{center}
\label{fig2} 
\vspace{-0.2cm} 
\begin{quote} {\footnotesize Figure 4: Stability diagram for synchronously updated CMLs 
with a von Neumann neighborhood of order 1. 
For selected values of $\gamma = 1-\beta$, the distribution of states
is plotted versus coupling strength $\epsilon$. Simulation details are as in Fig.~3.}     
\end{quote} 
\end{figure} 

\renewcommand{\baselinestretch}{0.85}
\begin{figure}[h] 
\begin{center} 
\epsfig{figure=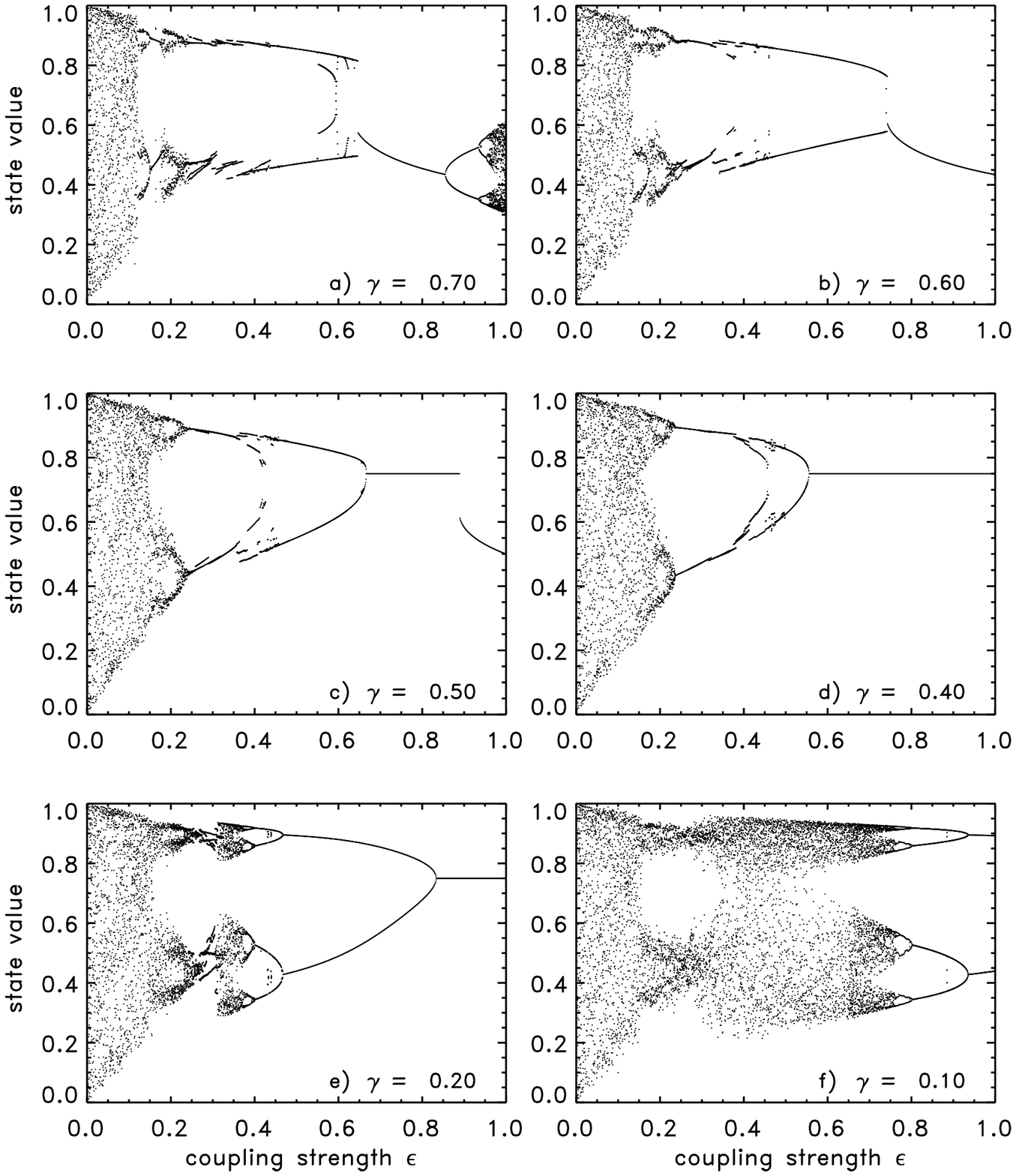,scale=0.7} 
\end{center}
\label{fig2} 
\vspace{-0.2cm}
\begin{quote} {\footnotesize Figure 5: Stability diagram for synchronously updated CMLs 
with a Moore neighborhood of order 1. 
For selected values of $\gamma = 1-\beta$, the distribution of states
is plotted versus coupling strength $\epsilon$. Simulation details are as in Fig.~3.}     
\end{quote} 
\end{figure} 

The stability conditions for $r=4$ read:
\[   1 < \epsilon \left( 2 + \lambda_k 
     (6\gamma-2) \right) < 3 ~. \]
For $\gamma = \frac{1}{3}$, we find $\epsilon_{\rm c}=\frac{1}{2}$
from the lower bound, while the upper bound gives no restriction.
For $\frac{1}{6} <\gamma<\frac{1}{3}$, only the maximal
eigenvalue is relevant and the conditions are independent
of the considered graph. We find:
\[    \epsilon > \frac{1}{6\gamma}  ~. \]
This is confirmed by the numerical results for a first order von Neumann
neighborhood in Fig.~4b-e and for a first order Moore neighborhood in Fig.~5e.
 
\vfill\eject

There is no stability for $\gamma< \frac{1}{6}$. (The
upper bound leads to no restriction.) This can be seen in Fig.~4e-f
for a first order von Neumann neighborhood and in Fig.~5f for 
a first order Moore neighborhood.

For $\gamma > \frac{1}{3}$ we find:
\[
\begin{array}{lcll}
   \mbox{v.\,N.\,1  } & & \D \frac{1}{4-6\gamma} < \epsilon 
          \hspace{1.5cm} & {\rm if}~~~ \D
              \frac{1}{3} < \gamma < \frac{1}{2} \\[0.3cm]
   \mbox{M.\,1  } & & \D \frac{1}{3-3\gamma} < \epsilon <
              \frac{1}{2\gamma}
     & {\rm if} ~~~ \D \frac{1}{3} < \gamma < \frac{3}{5}
             \\[0.3cm]  
   \mbox{v.\,N.\,2  } & & \D \frac{18}{49-39\gamma} <
           \epsilon < \frac{1}{2\gamma} & \D
      {\rm if}~~~ \frac{1}{3} < \gamma < \frac{49}{75} \\[0.3cm]
   \mbox{M.\,2  } & & \D \frac{48}{125-87\gamma} < \epsilon <
           \frac{1}{2\gamma} & \D {\rm if} ~~~
           \frac{1}{3} < \gamma < \frac{125}{183} \\[0.3cm]
   \mbox{global } & & \D \frac{1}{2} < \epsilon <
           \frac{1}{2\gamma} & \D {\rm for \ all } ~
           \gamma ~.
\end{array}  \]

For the regime $\frac{1}{6} < \gamma < \frac{1}{3}$,
the theoretical results are confirmed by the example in Fig.~4a for a first order
von Neumann neighborhood and in Fig.~5d for a first order Moore neighborhood.

While a first order von Neumann neighborhood produces no stable solution
for  $\gamma \ge 0.5$, this is different for a first order Moore neighborhood
(and for higher order neighborhoods). Figure 5c shows $\epsilon_c = 0.67$
at $\gamma = 0.5$ 
for a first order Moore neighborhood. In addition to the theoretically 
predicted stability of the unstable fixed point for all $\epsilon_c < 1$, anti-causal
coupling ($\gamma \neq 0$) yields
$\epsilon$-dependent stable solutions as well. For $r=4$ they are now given by 
\begin{equation}
\bar{u}_{1/2} = \frac{5}{8} \pm \frac{1}{8}
         \sqrt{ 9 - \frac{4}{\epsilon \gamma}}~. 
\end{equation}
Real solutions exist only for 
$\epsilon \geq \frac{4}{9 \gamma}$ and can be recognized in Fig.~5a-c. In particular,
Fig.~5c shows a discontinuous transition of the constant solution from
$\bar u = \frac{r-1}{r} = \frac{3}{4}$ to  $\bar u = \frac{5}{8}$ at $\epsilon = \frac{8}{9}$,
which represents a second critical coupling value at $\gamma = 0.5$ in Fig.~6.

The constant solution for a first order Moore neighborhood becomes unstable 
when $\gamma > \frac{1}{2}$ and $\epsilon < \frac{1}{2\gamma}$. This is confirmed in Fig.~5a-b. 
Upper and lower bounds coincide for $\gamma = 0.6$. 

\renewcommand{\baselinestretch}{0.85}
\begin{figure}[h] 
\begin{center} 
\epsfig{figure=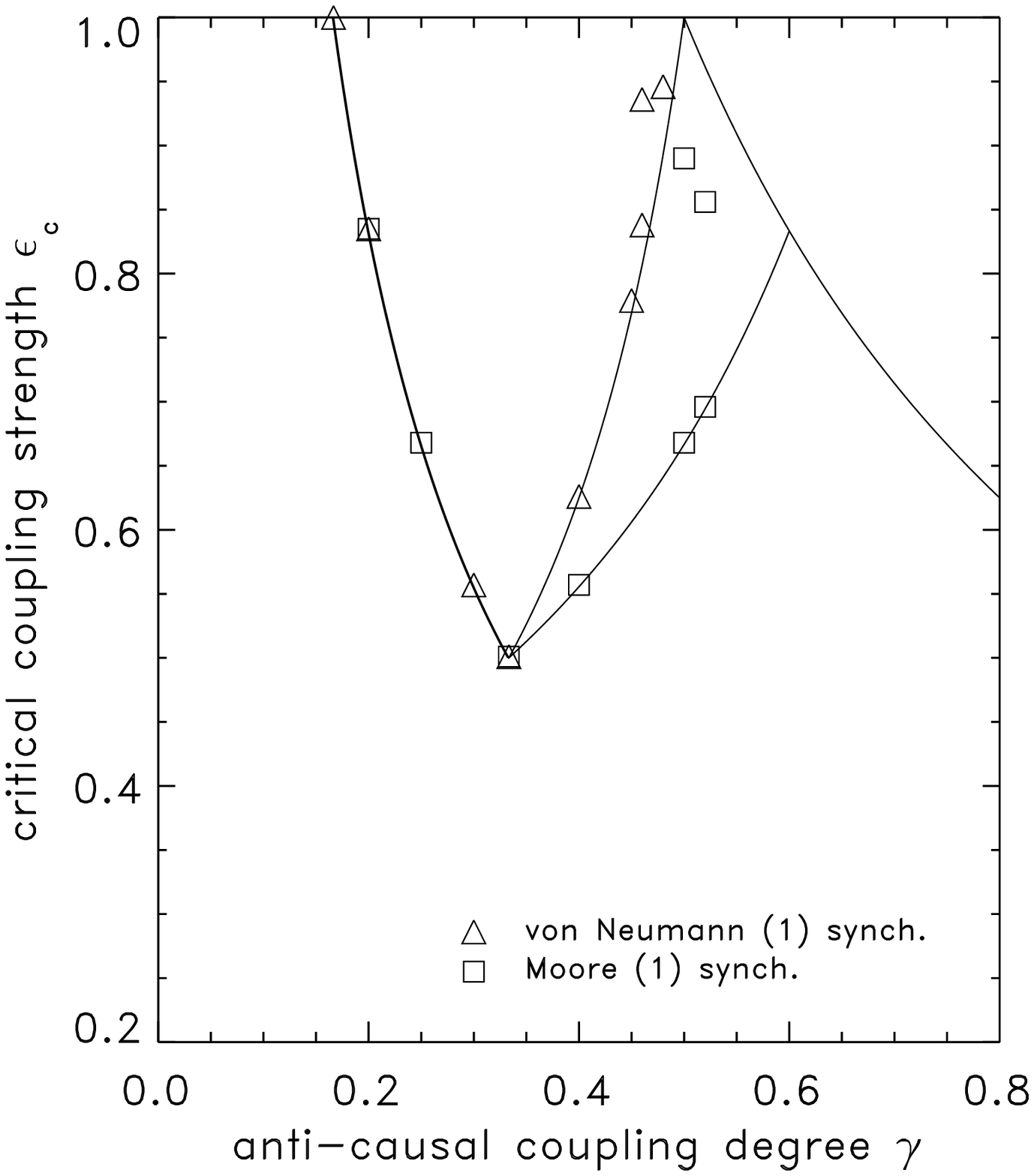,scale=0.5} 
\end{center}
\label{fig2} 
\vspace{-0.2cm}
\begin{quote} {\footnotesize Figure 6: Critical coupling strength $\epsilon_{c}$ 
for stabilization onset as a function of the degree $\gamma = 1-\beta$ of anti-causal 
versus non-causal coupling for $r = 4$ and for synchronous updating. The solid lines 
represent theoretical predictions and the dots refer to numerically obtained values
for von Neumann and Moore neighborhods of first order.}     
\end{quote} 
\end{figure} 

All results for combinations of non-causal and anti-causal coupling with synchronous
update for first order neighborhoods are compactly summarized in Figure 6, representing 
$\epsilon_c$ as a function of $\gamma = 1-\beta$ (for $r=4$). Most numerical data and theoretical 
curves agree perfectly. Results around $\gamma = 0.5$, deviating from the curves in Fig.~6,
do still satisfy the stability inequalities. Possible reasons for these deviations
are due to coexisting stable solutions. This is presently under investigation.

\section{Asynchronous Updating}
\label{sec:asynch}

Our results so far were derived for a synchronous updating scenario
for the CML configurations. For asynchronous updating 
the analytic treatment is more elaborate. The possibility of different 
types of asynchronous updates makes this even worse. 

For synchronous updating, we can interprete the CML (eq.\ \ref{eq:maincml})
as the iteration of a single mapping $F$ which maps a 
vector $\{u(n,x)\}$ (with components $x$) to a vector 
$\{u(n+1,x)\}$:
\[    F:  \{u(n,x)\}~ \longrightarrow ~
    \{u(n+1,x)\}   \]
where $u(n+1,x)$ is given by eq.\ (\ref{eq:maincml}) for
all $x$ (i.e., synchronously). The linear approximation around the constant solution
(leading to eq.\ \ref{eq:delta}) can be written as
\[    \delta u_{n+1}(x) = \sum_y F'(x,y) \delta u_n(y)\] 
where the matrix $F'(x,y)$ is given by:
\[   F'(x,y) = (1-\epsilon)f'(\bar{u}) \delta(x,y)
     + \epsilon g'(\bar{u}) (V^{-1}A)(x,y)   \]
with $\delta(x,y)$ as Kronecker's $\delta$.
The stability condition (eq.\ \ref{eq:stab1}) is simply
a condition for the eigenvalues of $F'$.

By contrast, asynchronous updating corresponds to a set of mappings 
(one for each vertex $z$) such that:
\[   F_z: \{ u(n,x) \}~ \longrightarrow ~
     \{ u(n+1,x)\}  ~, \]
where now $\{u(n+1,x)\}$ is given by:   
\[  u(n+1,x) = \left\{ \begin{array}{rl}
    u(n,x) & \mbox{for $x\neq z$} \\[0.2cm]
  {\displaystyle
 (1-\epsilon)f(u(n,x)) + \frac{\epsilon}{n_x}
   \sum_{y\sim x} g(u(n,x))} & \mbox{for $x=z$}
    \end{array} \right.  ~. \]
Correspondingly, the linear approximation now
depends on the vertex $z$, for which the update has
been made:
\[   F'_z(x,y) = \left\{  \begin{array}{rl}
            \delta(x,y) & \mbox{for $x\neq z$}\\
    F'(x,y) & \mbox{for $x=z$}  \end{array}
       \right.  ~. \]
The stability condition is now a condition for the
eigenvalues of a matrix, which is a product of 
$F'_z$-matrices,
where the product runs over $z$ either randomly
or according to some specific sequential rule. 
In general, the matrices $F'_z(x,y)$ do not commute 
for different
values of $z$ which makes it difficult to determine
the eigenvalues of this product.

However, despite this difficulty to derive an exact analytical
solution for the stability of asynchronously updated
configurations, there is an intuitive argument for the stability of
such configurations.

Consider some perturbation $\delta u(x)$ around a
constant solution $\bar{u}$ of the CML. The
lower bound of the stability
equation (eq.\ \ref{eq:stab1}) corresponds to a
change of $\delta u(x)$, including a
change in sign, at the vertex $x$ where the update is made. 
In particular, a perturbation proportional to the
eigenfunction of the adjacency matrix
for which the corresponding eigenvalue  
violates the lower stability bound for the
constant solution is simply multiplied by $-1$
at the critical value of $\epsilon$.

When an update
is made at some vertex $z$, we can assume that its
neighbors have been randomly updated before and
therefore their signs have changed randomly.
This implies that, on average, the
neighbors of vertex $z$ cancel and do not contribute to
the update, so we have an effective update equation of the form:
\[  \delta u(n+1,z) = (1-\epsilon) f'(\bar{u})
         \delta u(n,z) ~. \]
The lower bound of the stability condition 
of this equation reads:
\[   -1 < (1-\epsilon)f'(\bar{u})  ~, \]
which for the logistic mapping becomes:
\[   \epsilon > \frac{r-3}{r-2} ~. \]
This equation has been derived from different arguments
in \cite{atma04b}. It corresponds to the solid curve shown in Fig.~1. 

However, if $\epsilon$ approaches the upper bound of the 
stability inequality, a perturbation
$\delta u(x)$ does not change its sign. Therefore,
even if a random number of neighbors of a vertex
$z$, which is to be updated, have already been
updated before, their fluctuations will not
necessarily cancel on average. This situation 
is similar to synchronous updating. This argument
makes it plausible that the upper bound for $\epsilon$, 
as derived from synchronous updating, is also
valid for asynchronous updating.

Among the various situations studied numerically 
in \cite{atma04a, atma04b}, there is only one 
case where the upper bound of the stability inequality becomes relevant: 
combinations of causal and non-causal coupling in the regime $\alpha < \frac{1}{3}$
for a first order von Neumann neighborhood.
For this case the upper bound stability condition 
(\ref{eq:stabilnoncausal}) reads:
\[  \epsilon < \frac{3}{4-3\alpha}  ~. \]
This has been numerically confirmed for a number of examples shown in 
Fig.~2. 

\section{Generalized Neighborhoods} 
\label{sec:general}

In the preceding sections, all neighbors of a vertex have been treated
identically. However, it is possible to consider more general, inhomogeneous 
couplings of the vertices of the graph on which the CML is defined. 
For instance, the coupling strength may depend
on the distance on the graph (where distance referes to
the length of the minimal path). Such generalized couplings
are also suggested by renormalization methods, which, even
if the starting point is a nearest neighbor coupling, effectively
lead to couplings between more distant points. 

The most general case is given by a symmetric adjacency
matrix with a non-negative coupling between any two vertices:
\[   A^{(g)}(x,y) = A^{(g)}(y,x) \geq 0 \hspace{1cm}
     A^{(g)}(x,x)=0 ~. \]
We can define the generalized valence matrix by:
\[    V^{(g)}(x,y) = \left\{ \begin{array}{ll}
       \sum_z A^{(g)}(x,z) & \mbox{for } x=y \\
        0 & \mbox{otherwise} \end{array} \right. ~, \]
and we assume that $V^{(g)}(x,x)>0$ for all $x$. As in Sec.~2, 
one can show that $({\bf 1} \pm {V^{(g)}}^{-1}A^{(g)})$ 
is non-negative and that the largest eigenvalue of 
${V^{(g)}}^{-1}A^{(g)}$ is $+1$
(corresponding to the constant function on the graph).
In the context of Markov processes the matrix
${V^{(g)}}^{-1}A^{(g)}$ is called a Markov matrix. 

A simple example is given by a mixture of von Neumann 
and Moore neighborhoods of first order, where nearest horizontal and
vertical neighbors
are linked by a coupling $a_1$ and the additional diagonal 
neighbors are linked by a coupling $a_2$. We now define the 
generalized adjacency matrix 
\[  A(x,y) = \left\{ \begin{array}{ll}
    a_1 & \mbox{if $x$ and $y$ are nearest neighbors}\\
    a_2 & \mbox{if $x$ and $y$ are diagonal neighbors}\\
    0 & \mbox{otherwise}
    \end{array} \right.  ~. \]
This matrix can be written in the form:
\[  A(x,y) = a_1 (P_1 + P_1^{-1} + P_2 + P_2^{-2}) +
   a_2 (P_1P_2 + P_1 P_2^{-1} + P_1^{-1} P_2 + P_1^{-1} P_2^{-1})~,\]
and the corresponding eigenvalues $\hat\lambda$ of $A$ are:
\[  \hat\lambda = 2a_1 (\cos \phi_1 + \cos \phi_2) +
        4a_2 \cos \phi_1 \cos \phi_2 ~. \]
The maximal eigenvalue (equal to the generalized valence of
a vertex) and the minimal eigenvalue of $A$ are:
\[   \hat\lambda_{\rm max} = v= 4(a_1+a_2) \hspace{1cm}
    {\rm and} ~~~~ \hat\lambda_{\rm min} = - 4 {\rm max}(a_1,a_2)~. \]
The normalized minimal eigenvalue can now be tuned between
the case of first order von Neumann ($\lambda_{\rm min} = 
-1$) for $a_2=0$ (equivalently, $a_1=0$) and first order Moore  
($\lambda_{\rm min}=-\frac{1}{2}$) for $a_1=a_2$. 

An interesting case occurs if the coupling $a(L)$ 
depends only on the lattice-distance $L$ between two
points such that $a(L_1+L_2)=a(L_1)a(L_2)$. In this case
$a(L)=p^L$ for some positive constant $p$ which we assume to
be smaller than 1 in order to avoid divergencies on infinite
lattices. For the following we assume the lattice to be
rectangular, $d$-dimensional and consider the limit of
infinite lattice size, $N\rightarrow \infty$.
The adjacency matrix can then be written in the form:
\[   A = \prod _{i=1}^d 
     \left( {\bf 1} + \sum_{L=1}^\infty p^L (P_i^L+ P_i^{-L})
   \right) - {\bf 1} \]
with eigenvalues:
\[  \lambda = \prod_{i=1}^d 
   \left( 1 + 2 \sum_{L=1}^\infty p^L \cos L \phi_i
   \right) -1 
  = \prod_{i=1}^d  \left( \frac{1-p^2}{1-2p\cos \phi_i
          +p^2} \right)  -1 ~. \]
For the maximal and minimal normalized eigenvalues we finally obtain:
\[  \lambda_{\rm max} = 1
   \hspace{1cm}
    \lambda_{\rm min} = - \left( \frac{1-p}{1+p} \right)^d ~. \]

The significance of this exercise is obvious. Already for the
1-dimensional lattice (the circular graph) we can mimick, 
by a suitable tuning of $p$, the largest and smallest
eigenvalues of any other arbitrary graph or generalized
coupling. These eigenvalues are independent of the 
functions $f(x)$ (for which we used the logistic map) and $g(x)$.  
The stability of the constant solution of a
CML does not depend on any other property of the graph 
except these two eigenvalues. 
 
\section{Summary} 
\label{sec:summary}

Coupled map lattices (CMLs) and other complex model systems produce an 
overwhelming  phenomenological variety of features which are notoriously 
difficult to classify and explain in a compact and comprehensive way.
A promising concept for such purposes is the concept of stability
since it characterizes complex systems in their different dynamical regimes
in a most basic fashion.

The stability of CMLs has been studied numerically by a number of authors before.
A large class of such studies were carried out in the context of synchronizing and
controlling complex systems. In contrast, analytical tools explaining the
numerical results have not been broadly applied so far. Insofar as CMLs can be
considered as implemented on 
special types of graphs, spectral graph theory provides a particularly convenient
option to analyze the stability of CMLs theoretically. The maximal and minimal 
eigenvalues of the normalized adjacency matrix of the graph, on which the CML is defined, 
determine its stability properties.
 
We used these tools to analyze the stability properties of CMLs for 
different types of neighborhoods, different coupling scenarios and for different
updating rules. In this paper we focused on the stability at or around temporally constant
solutions of the CML as a whole. Particularly interesting is the case of globally stable 
solutions at locally unstable fixed points, which are independent of the coupling
strength $\epsilon$. In addition we discovered constant 
solutions which are explicitly $\epsilon$-dependent. 

For synchronous updating of the CML, we derived exact stability conditions for
causal, non-causal, and anti-causal coupling. These different coupling scenarios
distinguish whether states of a vertex of the graph are assumed to interact with 
their preceding, simultaneous, or future neighboring states, respectively. The considered 
neighborhoods are of low-order von Neumann and Moore type. All theoretical 
results agree excellently with numerical studies published earlier. Novel 
numerical simulations, referring to situations not treated before, additionally 
confirm our theoretical results.          

For asynchronous updating, the stability analysis is in general more involved
than for synchronous updating. In 
the case of a random updating sequence, the derivation simplifies and, again, 
gives results that agree with those of earlier numerical studies. A heuristic stability 
condition derived earlier could be confirmed as well. 

Finally, we generalized our approach to inhomogeneous neighborhoods, such as 
mixtures of neighborhoods of different types or coupling strenghts depending on 
the distance from a vertex. It could be shown that the stability of constant
solutions of CMLs is completely characterized by the maximal and minimal
eigenvalues of its normalized adjacency matrix, independent of the dynamics of the vertices
and independent of the causality features assumed for coupling. 

\section*{Acknowledgments}

We are grateful to one of the referees for several detailed and helpful suggestions
and corrections concerning the first version of the manuscript.

\bibliographystyle{unsrt}

\begin{thebibliography}{10}

\bibitem{turi52} A.~Turing.
\newblock The chemical basis of morphogenesis.
\newblock {\it Transactions of the Royal Society London, Series B} {\bf 237}, 37--72 (1952). 

\bibitem{kane93} K.~Kaneko, ed.
\newblock {\it Theory and Applications of Coupled Map Lattices}.
\newblock Wiley, New York, 1993. 

\bibitem{kuhn04} A.~Kuhn, A.~Aertsen, and S.~Rotter.
\newblock Neuronal integration of synaptic input in the fluctuation-driven regime.
\newblock {\it Journal of Neuroscience} {\bf 24}, 2345--2356 (2004). 

\bibitem{meht00} M.~Mehta and S.~Sinha.
\newblock Asynchronous updating of coupled maps leads to synchronization.
\newblock {\it CHAOS} {\bf 10}, 350--358 (2000). 

\bibitem{atma04a} H.~Atmanspacher and H.~Scheingraber. 
\newblock Stabilization of causally and non-causally coupled map lattices.
\newblock {\it Physica A} {\bf 345}, 435--447 (2005). 

\bibitem{maso03} C.~Masoller, A.C.~Marti, and D.H.~Zanette.
\newblock Synchronization in an array of globally coupled maps with delayed interactions.
\newblock {\it Physica A} {\bf 325}, 186--191 (2003).

\bibitem{liet04} C.~Li, S.~Li, X.~Liao, and J.~Yu.
\newblock Synchronization in coupled map lattices with small-world delayed interactions. 
\newblock {\it Physica A} {\bf 335}, 365--370 (2004).

\bibitem{atay04} F.~Atay, J.~Jost, and A.~Wende.
\newblock Delays, connection topology, and synchronization of coupled chaotic maps.
\newblock {\it Physical Review Letters} {\bf 92}, 144101 (2004).

\bibitem{atma04b} H.~Atmanspacher and  H.~Scheingraber.
\newblock Inherent global stabilization of unstable local behavior in coupled map lattices.
\newblock {\it Int.~Journal of Bifurcation and Chaos}, in press (2005).   
lanl preprints nlin.CD/0407008.   

\bibitem{atma05} H.~Atmanspacher, T.~Filk, and  H.~Scheingraber.
\newblock The significance of causally coupled, stable neuronal assemblies for the 
psychological time arrow.
\newblock Submitted.

\bibitem{kane00} K.~Kaneko and I.~Tsuda.
\newblock {\it Complex Systems: Chaos and Beyond}.
\newblock Springer, Berlin, 2000. 

\bibitem{lume94}  E.D.~Lumer and G.~Nicolis.
\newblock Synchronous versus asynchronous dynamics in spatially distributed systems.
\newblock {\it Physica D} {\bf 71}, 440--452 (1994).   

\bibitem{marc97} P.~Marcq, H.~Chat\'e, and P.~Manneville.
\newblock Universality in Ising-like phase transitions of lattices of coupled chaotic maps.
\newblock {\it Physical Review E} {\bf 55}, 2606--2627 (1997). 

\bibitem{rolf98} J.~Rolf, T.~Bohr, and M.H.~Jensen.
\newblock Directed percolation universality in asynchronous evolution of 
spatiotemporal intermittency.
\newblock {\it Physical Review E}  {\bf 57}, R2503--R2506 (1998). 

\bibitem{ott90} E.~Ott, C.~Grebogi, and J.A.~Yorke.
\newblock Controlling chaos. 
\newblock {\it Physical Review Letters} {\bf 64}, 1196--1199 (1990). 

\bibitem{mack95} M.~Mackey and J.~Milton.
\newblock Asymptotic stability of densities in coupled map lattices.
\newblock {\it Physica D} {\bf 80}, 1--17 (1995). 

\bibitem{loss95} J.~Losson, J.~Milton, and M.C.~Mackey. 
\newblock Phase transitions in networks of chaotic elements with short and long
range interactions.
\newblock {\it Physica D} {\bf 81}, 177--203 (1995).

\bibitem{lema96} A.~Lemaitre, H.~Chat\'e, and P.~Manneville.
\newblock Cluster expansion for collective behavior in discrete-space dynamical systems.
\newblock {\it Phys.~Rev.~Lett.} {\bf 77}, 486--489 (1996). 

\bibitem{lema98} A.~Lemaitre and H.~Chat\'e.
\newblock Nonperturbative renormalization group for chaotic coupled map lattices.
\newblock {\it Phys.~Rev.~Lett.} {\bf 80}, 5528--5531 (1998).

\bibitem{bely00} V.~Belykh, I.~Belykh, N.~Komrakow, and E.~Mosekilde.
\newblock Invariant manifolds and cluster synchronization in a family of locally
coupled map lattices.
\newblock {\it Discrete Dynamics} {\bf 4}, 245--256 (2000).

\bibitem{jost02} J.~Jost and M.P.~Joy.
\newblock Spectral properties and synchronization in coupled map lattices.
\newblock {\it Physical Review E} {\bf 65}, 016201 (2002). 

\bibitem{bely04} V.N.~Belykh, I.V.~Belykh, and M.~Hasler.
\newblock Connection graph stability method for synchronized coupled chaotic systems.
\newblock {\it Physica D} {\bf 195}, 159--187 (2004).

\bibitem{bigg74} N.~Biggs.
\newblock {\it Algebraic Graph Theory}.
\newblock Cambridge University Press, 1974.   

\bibitem{cvet95} D.M.~Cvetkovi\'c, M.~Doob, and H.~Sachs.
\newblock {\it Spectra of Graphs}.
\newblock Johann Ambrosius Barth, Heidelberg, 1995.



















\bibitem{wils85} R.J.~Wilson.
\newblock {\it Introduction to Graph Theory}.
\newblock Longman Scientific \& Technical, Essex, 1985.

\end{thebibliography}
 
\end{document}